\documentclass[a4paper,fleqn]{cas-sc}

\usepackage[authoryear,longnamesfirst]{natbib}
\usepackage{array}
\usepackage{booktabs}

\graphicspath{{thumbnails}/}

\begin{document}
\let\WriteBookmarks\relax
\def\floatpagepagefraction{1}
\def\textpagefraction{.001}

\shorttitle{Not-Quite-Transcendental Functions}    

\shortauthors{Miller, Dolence, and Holladay}  

\title[mode = title]{Not-Quite-Transcendental Functions and their Applications}

\author[1,2]{Jonah M. Miller}[orcid=0000-0001-6432-7860]
\fnmark[1] 
\ead{jonahm@lanl.gov}
\author[1,2]{Joshua C. Dolence}
\fnmark[2] 
\ead{jdolence@lanl.gov} 
\author[3]{Daniel Holladay}
\fnmark[3]
\ead{danl@lanl.gov}

\affiliation[1]{
  organization={Computational Physics and Methods, Los Alamos National Laboratory},
  addressline={P.O. Box 1663},
  city={Los Alamos},
  state={NM},
  country={USA}}
\affiliation[2]{
  organization={Center for Theoretical Astrophysics, Los Alamos National Laboratory},
  addressline={P.O. Box 1663},
  city={Los Alamos},
  state={NM},
  country={USA}}
\affiliation[3]{
  organization={Applied Computer Science, Los Alamos National Laboratory},
  addressline={P.O. Box 1663},
  city={Los Alamos},
  state={NM},
  country={USA}}

\begin{abstract}[ S U M M A R Y]

  Transcendental functions, such as exponentials and logarithms,
  appear in a broad array of computational domains: from simulations
  in curvilinear coordinates, to interpolation, to machine
  learning. Unfortunately they are typically expensive to compute
  accurately. In this note, we argue that in many cases, the
  \textit{properties} of the function matters more than the exact
  functional form. We present new functions, which are not
  transcendental, that can be used as drop-in replacements for the
  exponential and logarithm in many settings for a significant
  performance boost. We show that for certain applications using these
  functions result in no drop in the accuracy at all, as they are
  perfectly accurate representations of themselves, if not the
  original transcendental functions.

\end{abstract}




\maketitle

\section{Introduction}
\label{sec:intro}

A transcendental function cannot be expressed as a finite series of
algebraic operations---in other words, it transcends algebra
\citep{townsend1915functions}. The implementation of a transcendental
function on a computer, thus, typically relies on a computationally
expensive finite truncation of an infinite series, with sufficiently
many terms chosen to satisfy a required precision
\citep{cody1980software}. Unfortunately, transcendental functions,
such as exponentials, logarithms, and trigonometric functions, are
ubiquitous in mathematics, science, and engineering, making them a
necessary evil in computational science. This ubiquity has lead to a
plethora of approaches in both hardware and software for accelerating
the approximate calculation of transcendental functions stretching
back decades \citep{Marino1672109, PERINI201837}. Approximations
carry with them a cost in accuracy, which must be accounted for.

However, there are several settings where transcendental functions are
used where what matters is the \textit{properties} of the function,
rather than its precise, transcendental nature. We therefore propose
that, in these settings, one may replace certain transcendental
functions with non-transcendental functions that look like the
original functions but are easier to compute. We call these functions
\textit{not-quite transcendental} (NQT). NQT functions are
\textit{not} necessarily approximations to transcedental functions,
although they may be (and have broadly been) applied that way. Rather
they are \textit{exact} functional representations of themselves that
share desirable properties of a given transcendental function.

In the remainder of this note, we cover several case studies where NQT
functions may be applied with \textit{no loss of accuracy} instead of
the traditionally used transcendental function: coordinate
transformations, logarithmic interpolation, activation functions in
neural networks; and maximum likelihood estimation. We use the first
case study to introduce the NQT functions we will
study---exponentials, logarithms, and hyperbolic functions. After the
case studies we present performance and accuracy results, before
offering a few concluding thoughts.

\section{First Case Study: Coordinate Transformations}
\label{sec:transformations}

Consider a problem with large dynamic range. For example, in
simulations of the in-spiral and merger of two neutron stars, as
observed by the LIGO-VIRGO gravitational wave interferometers
\citep{GW170817}, dynamics occur at a length scale of a few tens of
kilometers, but gravitational waves must be extracted from a
simulation in the far zone, perhaps thousands of kilometers away from
the binary. Models of the accretion disk around the supermassive black
hole at the center of our galaxy---as recently dramatically observed
by the Event Horizon Telescope \citep{EHTSagA}---suffer a similar
difficulty. Dynamics happens on relatively short length scales, near
the event horizon, but boundary conditions are only known far from the
black hole. This difficulty appears commonly in astrophysics but
extends to applications as far ranging as quantum chemistry; The
primary modes of the radial component of electron wavefunctions in an
atom occur at roughly a Bohr radius, however, the wavefunctions are
normalized via an integral extending over all space
\citep{grant2007relativistic}.

One way of capturing this dynamic range is through adaptive mesh
refinement \citep{BERGER1984484} or a non-uniform grid. Another
approach, however, is to apply a coordinate transformation. A
spherical simulation grid with grid spacing growing exponentially in
radius may be implemented by defining a new coordinate,
\begin{equation}
  \label{eq:log:transform}
  \tilde{r} = \ln(r)
\end{equation}
and writing the governing equations in terms of $\tilde{r}$ rather
than $r$. This is exactly the approach taken in the vast majority of
simulation codes used by the Event Horizon Telescope team
\citep{McKinney_2004}. One may even bring the entire real number line
to finite grid coordinates by using a transformation such as
$\tilde{x} = \tanh(x)$ so that a physically infinite distance has a
coordinate value of $\pm 1$. Such was the approach taken by the first
simulation of a full orbit of a binary black hole system
\citep{Pretorius}. However, what matters in equation
\eqref{eq:log:transform} is not that a logarithm is used. Rather the
important property of the transformation is that it is
\textit{approximately} logarithmic---that derivatives of the
transformation scale roughly inversely with $r$, so that the physical
grid spacing roughly scales with $r$. We now introduce the NQT
function that meets these needs.

\section{The Not-Quite-Transcendental Functions for Logarithms and Exponentials}
\label{sec:introducing:NQT}

We begin with the well-known exploit of the construction of a floating
point number. A positive floating point number $n$ is internally
represented as
  $n = m \times 2^p,$
where $m\in (0, 1/2)$ is the \textit{mantissa} and the integer $p$ is
the \textit{exponent}. Most programming languages and hardware vendors
provide the ability to pull apart $n$ into its components, $m$ and
$p$.  This implies that
\begin{equation}
  \label{eq:approx:log}
  \lg(n) = \lg(m) + p
\end{equation}
which reduces the standard problem of computing a logarithm to
computing $\lg(m)$ on the interval $[0, 1/2)$. Change of basis
formulae can then move from $\lg$ to whatever logarithmic basis is
appropriate.

We \textit{define} the NQT function
\begin{equation}
  \label{eq:def:lg:NQT}
  \lg_{\text{NQT}}(n) = 2 (m - 1) + p,
\end{equation}
which is both continuous and differentiable (i.e., $C^1$). As observed
in \citet{Hall}, this function can be interpreted as a
piecewise-linear interpolation of $\lg(x)$, however we treat it as an
independent function in its own right. Moreover, it is exactly
invertible by simply finding $m$ and $p$ and computing $m \times
2^p$. We define this inverse function as
$\text{pow2}_{\text{NQT}}(x)$. Multiplication by a constant value is
all that is required to convert these functions to NQT versions of
$\log_{10}(x)$ and $10^x$. One can further optimize the implementation
of these NQT functions via so-called \textit{integer aliasing}, as
described in, e.g., \citet{Blinn}.

Hyperbolic functions and their inverses are composed of sums and
products of exponentials and logarithms, and so NQT hyperbolic
functions can be constructed by sums and products of NQT logs and
exponents. NQT hyperbolic tangent, for example, is given by
  $\tanh_{\text{NQT}}(x) = \left((e_{\text{NQT}}^{2x} - 1\right)/\left(e_{\text{NQT}}^{2 x} + 1\right),$
where $e_{\text{NQT}}^x$ is the NQT counterpart of the natural
exponential function.

$\lg_{\text{NQT}}$ and $\text{pow2}_{\text{NQT}}$ are
\textit{compatible} with commonly used low-accuracy approximations of
$\lg(x)$ and $2^x$---indeed they coincide with their transcendental
counterparts to a few percent. However we emphasize that they are
\textit{not} approximations in this context. If we deliberately define
a coordinate transformation given by
$\tilde{r} = \lg_{\text{NQT}}(r)$, that transformation is exact and
exactly invertible. \textit{No accuracy} is lost
in a simulation that uses NQT log coordinates compared to a simulation
that uses truly logarithmic coordinates.\footnote{We note that because these
  functions are only $C^1$ they may not be appropriate for coordinate
  transformations if a very high-order numerical method, such as a
  spectral method is used, as smoother (in the formal sense) functions
  may be required for good convergence.}

\section{Second Case Study: Logarithmic Interpolation}
\label{sec:logarithmic:interpolation}

Tabulated data often spans many orders of magnitude. Nuclear reaction
rates used in astrophysics are infamous in this regard
\citep{Reaclib}. Equations of state for real materials such as iron and
steel are another common example \citep{Sesame}. A common approach is
to apply a logarithmic coordinate transformation, as described in
Section \ref{sec:transformations}, to the independent and/or dependent
variable. This results in log-log or log-linear interpolation.

As with Section \ref{sec:transformations} above, the key is not that
the interpolation grid be exactly logarithmic. Rather the defining
trait is that the transformed interpolation grid can cover many orders
of magnitude such that the relative interpolation error remains
roughly constant. And as with Section \ref{sec:transformations} the
NQT counterparts to the logarithm and exponential functions satisfy
these criteria exactly---often better than their transcendental
counterparts.

\section{Third Case Study: Activation Functions}
\label{sec:activation}

A neural network is built on two core components: a linear operation
typically implemented as matrix multiplication, and a nonlinear
operation, called an \textit{activation} function
\citep{goodfellow2016deep}. Common activation functions include the
rectified linear unit (RELU) and the hyperbolic tangent. The important
properties of an activation function are that it is: (a) nonlinear and
(b) differentiable.

In the case of the hyperbolic tangent activation function, the exact
functional form matters less than the above key points, as well as the
ability to span the range $(-1, 1)$ with a non-vanishing
derivative. Activation functions are also ideally easy to implement
and cheap to compute. The NQT tangent function meets these criteria.

\section{Fourth Case Study: Maximum Likelihood Estimation}
\label{sec:MLE}

In probability theory one often wishes to compute the outcome that is
most probable. This procedure is called \textit{maximum likelihood
  estimation.} For some set of $N$ independent random variables
$x_1,\ldots x_N$ with probability density functions
$p_1(x_1)\ldots p_N(x_1)$, this procedure amounts to finding the
vector $(x_1, \ldots x_N)$ that maximizes the product
$\prod_{i=1}^N p_i(x_i)$. Most probability density functions are
logarithmically concave, making it convenient to transform this
optimization problem to maximize \citep{millar2011maximum}
$\ln(\prod_{i=1}^N p_i(x_i))$.  Here the important property of $\ln$
is that it is monotone and approximately logarithmic, which the NQT
log is.

Traditionally these calculations leverage the transitive property of
logs that $\ln(x y) = \ln(x) + \ln(y)$. Unfortunately, our NQT
functions are only approximately transitive, and so this
transformation can not necessarily be applied. That said, for MLE
estimation, if the argmax of $\sum_i \ln_{NQT}p(x_i)$ is the same as
the argmax of $\ln_{NQT}(\prod p(x_i))$, then one can leverage
transitivity.

\section{Performance Experiments}
\label{sec:results}

To investigate the performance of our NQT functions, we compare them
to the standard library (STL) implementations of their transcendental
counterparts in Table 1. We examine three relevant functions:
$f(x) = 10^x$, $f(x) = \log_{10}(x)$, and $f(x) = \tanh(x)$. Timings
in Table 1 are shown in nanoseconds per point, where we use a
sufficiently large grid to saturate the target hardware. We run on an
Nvidia V100 GPU as well as an Intel Xeon Gold 6254 CPU. In the latter
case, we ran our tests with both GNU and Intel compilers.

Broadly we find significant speedups across the board, with the
largest speedup being more than a factor of 8x for the exponential on
Cascade lake with the intel compiler. We find the least impressive
speedups for $\tanh(x)$, perhaps because it has received vendor
attention due to application in machine learning. Nevertheless,
speedups are at least 1.2x on the V100 and much larger on the Skylake.

\begin{table}
  \label{tab:results}
  \caption{Performance Comparison Between standard library (STL)
    implementations of transcendental functions and our
    implementations of their not-quite-transcendental (NQT)
    counterparts. Timings are in nanoseconds per point, while speedups
    are the performance improvement of the NQT implementation over the
    STL implementation. Calculation was performed on a sufficiently
    large vector to saturate the hardware.The V100 calculation was
    performed with cuda 11.4.2 with the GCC backend using GCC
    9.4.0. The cascade lake calculation was performed with either GCC
    9.4.0 or with the Intel 19.0.5 compiler as annotated. The CPU was
    an Intel Xeon Gold 6254.}
  \begin{tabular*}{\tblwidth}{l | c | c | c| c | c | c | c | c | c}
    \toprule
    {} &\multicolumn{3}{c}{$10^x$}&\multicolumn{3}{c}{$\log_{10}(x)$}&\multicolumn{3}{c}{$\tanh(x)$}\\
    \midrule
    \textbf{Architecture} & \textbf{STL} & \textbf{NQT} & \textbf{Speedup} & \textbf{STL} & \textbf{NQT} & \textbf{Speedup} & \textbf{STL} & \textbf{NQT} & \textbf{Speedup}\\
    \midrule
    Volta V100
       & 2.7e-2 & 5.7e-3 & 4.7 & 1.2e-2 & 5.7e-3 & 2.0 & 7.8e-3 & 6.3e-3 & 1.2\\
    \midrule
    Cascade Lake (gnu)
       & 5.8e+1 & 1.1e+1 & 5.1 & 3.1e+1 & 1.1e+4 & 2.7 & 3.4e+1 & 1.4e+1 & 2.2\\
    \midrule
    Cascade Lake (int)
       & 1.2e+1 & 1.4e+0 & 8.2 & 6.5e+0 & 1.3e+0 & 4.8 & 8.2e+0 & 1.9e+1 & 4.2\\
    \bottomrule
  \end{tabular*}
\end{table}

\section{Concluding Thoughts}
\label{sec:concluding:thoughts}

In this short note, we introduce the \textit{not-quite-transcendental}
counterparts to exponential, logarithmic, and hyperbolic
transcendental functions. We show that for several common problems in
computational science, transcendental functions are used because of
several desirable properties, but the exact form of the function
\textit{does not matter}. Therefore, using the
not-quite-transcendental counterpart offers a performance improvement
for \textit{no} accuracy cost. In our numerical experiments we find
perfect invertability of our NQT functions and a speedup over their
transcendental counterparts as high as 8.2x.

In future work, it would be interesting to find
not-quite-transcendental counterparts to trigonometric
functions. However, we expect that trigonomentric functions are more
often required exactly and thus fewer applications for
not-quite-transcendental versions would be available.

\section{Acknowledgements}
\label{sec:acknowledgements}

The authors thank Luke Roberts, Erik Schnetter, and Raymond Wong for
useful discussions. This work was supported by the U.S. Department of
Energy through the Los Alamos National Laboratory (LANL). LANL is
operated by Triad National Security, LLC, for the National Nuclear
Security Administration of U.S. Department of Energy (Contract
No. 89233218CNA000001). We used the Darwin testbed, which is funded by
the Computational Systems and Software Environments subprogram of
LANL's Advanced Simulation and Computing program (NNSA/DOE).


%
%

\bibliography{fast-logs}

\begin{thebibliography}{16}
\expandafter\ifx\csname natexlab\endcsname\relax\def\natexlab#1{#1}\fi
\providecommand{\url}[1]{\texttt{#1}}
\providecommand{\href}[2]{#2}
\providecommand{\path}[1]{#1}
\providecommand{\DOIprefix}{doi:}
\providecommand{\ArXivprefix}{arXiv:}
\providecommand{\URLprefix}{URL: }
\providecommand{\Pubmedprefix}{pmid:}
\providecommand{\doi}[1]{\href{http://dx.doi.org/#1}{\path{#1}}}
\providecommand{\Pubmed}[1]{\href{pmid:#1}{\path{#1}}}
\providecommand{\bibinfo}[2]{#2}
\ifx\xfnm\relax \def\xfnm[#1]{\unskip,\space#1}\fi
\bibitem[{Abbott et~al.(2017)}]{GW170817}
\bibinfo{author}{Abbott, B.P.}, et~al. (\bibinfo{collaboration}{LIGO Scientific
  Collaboration and Virgo Collaboration}), \bibinfo{year}{2017}.
\newblock \bibinfo{title}{Gw170817: Observation of gravitational waves from a
  binary neutron star inspiral}.
\newblock \bibinfo{journal}{Phys. Rev. Lett.} \bibinfo{volume}{119},
  \bibinfo{pages}{161101}.
\newblock \URLprefix
  \url{https://link.aps.org/doi/10.1103/PhysRevLett.119.161101},
  \DOIprefix\doi{10.1103/PhysRevLett.119.161101}.
\bibitem[{{Akiyama} et~al.(2022)}]{EHTSagA}
\bibinfo{author}{{Akiyama}, K.}, et~al., \bibinfo{year}{2022}.
\newblock \bibinfo{title}{{First Sagittarius A* Event Horizon Telescope
  Results. I. The Shadow of the Supermassive Black Hole in the Center of the
  Milky Way}}.
\newblock \bibinfo{journal}{Astrophysical Journal Letters}
  \bibinfo{volume}{930}, \bibinfo{pages}{L12}.
\newblock \DOIprefix\doi{10.3847/2041-8213/ac6674}.
\bibitem[{Berger and Oliger(1984)}]{BERGER1984484}
\bibinfo{author}{Berger, M.J.}, \bibinfo{author}{Oliger, J.},
  \bibinfo{year}{1984}.
\newblock \bibinfo{title}{Adaptive mesh refinement for hyperbolic partial
  differential equations}.
\newblock \bibinfo{journal}{Journal of Computational Physics}
  \bibinfo{volume}{53}, \bibinfo{pages}{484--512}.
\newblock \URLprefix
  \url{https://www.sciencedirect.com/science/article/pii/0021999184900731},
  \DOIprefix\doi{https://doi.org/10.1016/0021-9991(84)90073-1}.
\bibitem[{Blinn(1997)}]{Blinn}
\bibinfo{author}{Blinn, J.}, \bibinfo{year}{1997}.
\newblock \bibinfo{title}{Floating-point tricks}.
\newblock \bibinfo{journal}{IEEE Computer Graphics and Applications}
  \bibinfo{volume}{17}, \bibinfo{pages}{80--84}.
\newblock \DOIprefix\doi{10.1109/38.595279}.
\bibitem[{Cody and Waite(1980)}]{cody1980software}
\bibinfo{author}{Cody, W.}, \bibinfo{author}{Waite, W.}, \bibinfo{year}{1980}.
\newblock \bibinfo{title}{Software Manual for the Elementary Functions}.
\newblock Prentice-Hall Foundations of Earth Science Series,
  \bibinfo{publisher}{Prentice-Hall}.
\newblock \URLprefix \url{https://books.google.com/books?id=d4UZAQAAIAAJ}.
\bibitem[{Goodfellow et~al.(2016)Goodfellow, Bengio and
  Courville}]{goodfellow2016deep}
\bibinfo{author}{Goodfellow, I.}, \bibinfo{author}{Bengio, Y.},
  \bibinfo{author}{Courville, A.}, \bibinfo{year}{2016}.
\newblock \bibinfo{title}{Deep Learning}.
\newblock Adaptive Computation and Machine Learning series,
  \bibinfo{publisher}{MIT Press}.
\newblock \URLprefix \url{https://books.google.com/books?id=omivDQAAQBAJ}.
\bibitem[{Grant(2007)}]{grant2007relativistic}
\bibinfo{author}{Grant, I.}, \bibinfo{year}{2007}.
\newblock \bibinfo{title}{Relativistic Quantum Theory of Atoms and Molecules:
  Theory and Computation}.
\newblock Springer Series on Atomic, Optical, and Plasma Physics,
  \bibinfo{publisher}{Springer New York}.
\newblock \URLprefix \url{https://books.google.com/books?id=yYFCAAAAQBAJ}.
\bibitem[{Hall et~al.(1970)Hall, Lynch and Dwyer}]{Hall}
\bibinfo{author}{Hall, E.}, \bibinfo{author}{Lynch, D.},
  \bibinfo{author}{Dwyer, S.}, \bibinfo{year}{1970}.
\newblock \bibinfo{title}{Generation of products and quotients using
  approximate binary logarithms for digital filtering applications}.
\newblock \bibinfo{journal}{IEEE Transactions on Computers}
  \bibinfo{volume}{C-19}, \bibinfo{pages}{97--105}.
\newblock \DOIprefix\doi{10.1109/T-C.1970.222874}.
\bibitem[{Marino(1972)}]{Marino1672109}
\bibinfo{author}{Marino, D.}, \bibinfo{year}{1972}.
\newblock \bibinfo{title}{New algorithms for the approximate evaluation in
  hardware of binary logarithms and elementary functions}.
\newblock \bibinfo{journal}{IEEE Transactions on Computers}
  \bibinfo{volume}{C-21}, \bibinfo{pages}{1416--1421}.
\newblock \DOIprefix\doi{10.1109/T-C.1972.223516}.
\bibitem[{McKinney and Gammie(2004)}]{McKinney_2004}
\bibinfo{author}{McKinney, J.C.}, \bibinfo{author}{Gammie, C.F.},
  \bibinfo{year}{2004}.
\newblock \bibinfo{title}{A measurement of the electromagnetic luminosity of a
  kerr black hole}.
\newblock \bibinfo{journal}{The Astrophysical Journal} \bibinfo{volume}{611},
  \bibinfo{pages}{977--995}.
\newblock \URLprefix \url{https://doi.org/10.1086/422244},
  \DOIprefix\doi{10.1086/422244}.
\bibitem[{{Meisel}(2008)}]{Reaclib}
\bibinfo{author}{{Meisel}, Z.}, \bibinfo{year}{2008}.
\newblock \bibinfo{title}{{REACLIB: A Reaction Rate Library for the Era of
  Collaborative Science}}, in: \bibinfo{booktitle}{APS Division of Nuclear
  Physics Meeting Abstracts}, p. \bibinfo{pages}{DA.064}.
\bibitem[{Millar(2011)}]{millar2011maximum}
\bibinfo{author}{Millar, R.}, \bibinfo{year}{2011}.
\newblock \bibinfo{title}{Maximum Likelihood Estimation and Inference: With
  Examples in R, SAS and ADMB}.
\newblock Statistics in Practice, \bibinfo{publisher}{Wiley}.
\newblock \URLprefix \url{https://books.google.com/books?id=qvzUGYmChxUC}.
\bibitem[{P.(1992)}]{Sesame}
\bibinfo{author}{P., L.S.}, \bibinfo{year}{1992}.
\newblock \bibinfo{title}{Sesame : The los alamos national laboratory equation
  of state database}.
\newblock \bibinfo{journal}{Los Alamos National Laboratory report
  LA-UR-92-3407} \URLprefix
  \url{https://cir.nii.ac.jp/crid/1572824499059761024}.
\bibitem[{Perini and Reitz(2018)}]{PERINI201837}
\bibinfo{author}{Perini, F.}, \bibinfo{author}{Reitz, R.D.},
  \bibinfo{year}{2018}.
\newblock \bibinfo{title}{Fast approximations of exponential and logarithm
  functions combined with efficient storage/retrieval for combustion kinetics
  calculations}.
\newblock \bibinfo{journal}{Combustion and Flame} \bibinfo{volume}{194},
  \bibinfo{pages}{37--51}.
\newblock \URLprefix
  \url{https://www.sciencedirect.com/science/article/pii/S0010218018301652},
  \DOIprefix\doi{https://doi.org/10.1016/j.combustflame.2018.04.013}.
\bibitem[{Pretorius(2005)}]{Pretorius}
\bibinfo{author}{Pretorius, F.}, \bibinfo{year}{2005}.
\newblock \bibinfo{title}{Evolution of binary black-hole spacetimes}.
\newblock \bibinfo{journal}{Phys. Rev. Lett.} \bibinfo{volume}{95},
  \bibinfo{pages}{121101}.
\newblock \URLprefix
  \url{https://link.aps.org/doi/10.1103/PhysRevLett.95.121101},
  \DOIprefix\doi{10.1103/PhysRevLett.95.121101}.
\bibitem[{Townsend(1915)}]{townsend1915functions}
\bibinfo{author}{Townsend, E.}, \bibinfo{year}{1915}.
\newblock \bibinfo{title}{Functions of a Complex Variable}.
\newblock American mathematical series, \bibinfo{publisher}{H. Holt}.
\newblock \URLprefix \url{https://books.google.com/books?id=JDUNAAAAYAAJ}.

\end{thebibliography}
\bibliographystyle{cas-model2-names}

\end{document}